\RequirePackage{fix-cm}
\documentclass[smallextended]{svjour3}       
\smartqed  
\usepackage{braket}
\usepackage{pgfplots}
\usepackage{csquotes}
\usepackage{float}
\usepackage{multirow}
\usepackage{dcolumn}
\usepackage[]{hyperref}
\usepackage{hhline}
\usepackage{amssymb}
\usepackage{subfigure}
\usepackage{xcolor}
\usepackage{graphicx}
\usepackage{bm}
\usepackage{amsmath}
\usepackage{bbm}
\begin{document}

\title{Investigation of quantum pigeonhole effect in IBM quantum computer}

\author{Narendra N. Hegade \and
        Antariksha Das \and Swarnadeep Seth \and Prasanta K. Panigrahi }


\institute{Narendra N. Hegade \at
              Department of Physics, National Institute of Technology Silchar, Silchar 788010, India\\
            \email{narendrahegade5@gmail.com}           
           \and
           Antariksha Das \at
              QuTech, Delft University of Technology, Lorentzweg 1, 2628 CJ Delft, The Netherlands \\     
              \email{a.das-1@tudelft.nl} 
              \and
              Swarnadeep Seth \at   Department of Physics, University of Central Florida, Orlando, FL 32186, USA \\ 
              \email{swarnadeep@knights.ucf.edu}
              \and 
              Prasanta K. Panigrahi \at   Department of Physical Sciences, Indian Institute of Science Education and Research Kolkata, Mohanpur 741246, West Bengal, India \\ \email{pprasanta@iiserkol.ac.in}}

\date{Received: date / Accepted: date}

\maketitle

\begin{abstract}
Quantum pigeonhole principle states that if there are three pigeons and two boxes then there are instances where no two pigeons are in the same box which seems to defy classical pigeonhole counting principle. Here, we investigate the quantum pigeonhole effect on the ibmqx2 superconducting chip with five physical qubits. We also observe the same effect in a proposed non-local circuit which avoid any direct physical interactions between the qubits which may lead to some unknown local effects. We use the standard quantum gate operations and measurement to construct the required quantum circuits on IBM quantum experience platform. We perform the experiment and simulation which illustrates the fact that no two qubits (pigeons) are in the same quantum state (boxes). The experimental results obtained using IBM quantum computer are in good agreement with theoretical predictions.

\end{abstract}


\section{Introduction \label{I}}

Quantum mechanics is well known for its counter-intuitive results which pose a conceptual conflict of our regular understanding. There are many quantum mechanical phenomena such as the EPR paradox, the no-cloning theorem, quantum Zeno effect, quantum teleportation, quantum tunneling etc. that can not be answered by classical physics. Quantum pigeonhole effect is one of these. In number theory, classical pigeonhole principle \cite{Allenby2011Howtocount} states that if $n$ objects are distributed among $m$ boxes, with the condition $m < n$, then there is at least one box where we can find more than one object. In other words, if there are more objects than the number of boxes then there is at least one box which must contain more than one object. 

\begin{figure}
    \centering
    \includegraphics [width=\linewidth]{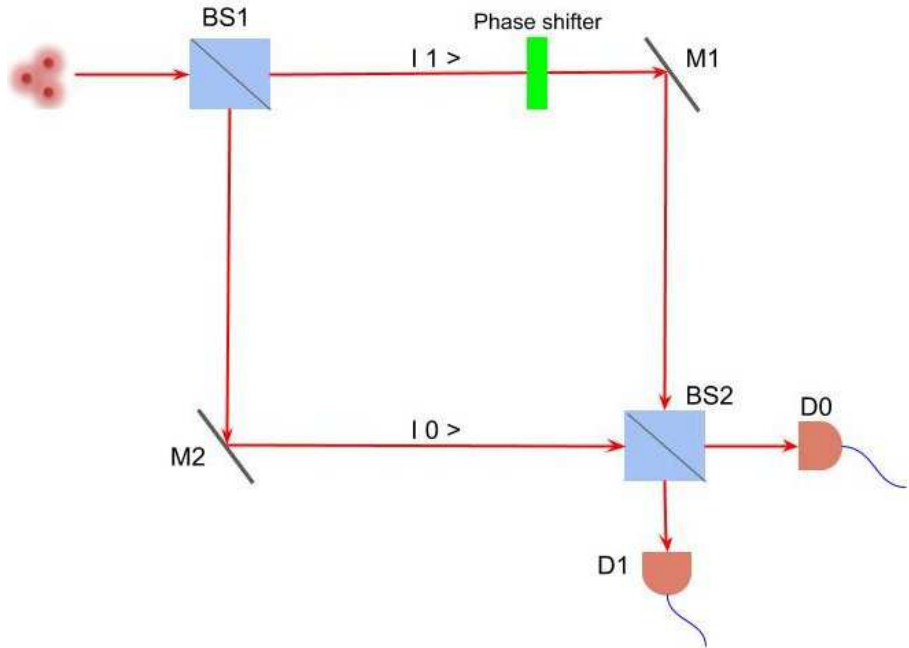}
    \caption{\textbf{Schematic diagram of the Mach-Zehnder interferometer}. Three quantum particles are injected simultaneously. They are split into the two arms of the interferometer ($\Ket{0}$ and $\Ket{1}$) after the first beam splitter $BS1$. There is a phase shifter in one path of the interferometer. The particles are detected at detector $D_0$ and $D_1$ after another beam splitter $BS2$.}
    \label{fig1}
\end{figure}

Aharonov et al. \cite{Aharonov2016QPP} first proposed the idea of quantum pigeonhole effect(QPHE) where they have shown that in some scenarios the classical pigeonhole counting principle is violated. It is shown that for a particular choice of pre- and post-selected state, three quantum particles which can take two quantum states could end up in a situation where no two particles can be found in the same quantum states. To observe the quantum pigeonhole effect, the former designed and performed an interferometric experiment shown in Figure \ref{fig1}. In their set-up three quantum particles (pigeons) pass simultaneously through the two arms (Pigeonholes/boxes) of the Mach-Zehnder interferometer (MZI) which characterizes two distinct quantum state ($\Ket{0}$ and $\Ket{1}$) of the quantum particles. Now because of the Coulomb repulsion between the quantum particles if at least two or three particles are in the same arm of the interferometer then the particles will repel each other and expected to get deflected and the pattern of the detector will give information whether any of the quantum particles are in the same path of the interferometer or not. The particles have equal probability of arriving at either of the two detectors. It is shown that if one post-select those cases where all three quantum particles are detected at the same detector then the pattern of the detector indicates that none of the quantum particles get deflected thus no such interaction has taken place. It suggests that no two quantum particles can take the same path which contradict the classical pigeonhole principle. This phenomenon has already drawn a fair amount of attention. Recently, Chen et al. \cite{chen2019ExpParadox} experimentally demonstrated the quantum pigeonhole paradox using three single photons, transmitting through two distinct polarization channels under appropriate pre and post selections of the polarization states. They used the weak measurement technique \cite{weakmeasure1,weakmeasure2,weakmeasure3} to probe the underlying mechanism. Mahesh et al. \cite{mahesh2016NMR} experimentally simulated the quantum pigeonhole principle using four-qubit NMR quantum simulator where the quantum pigeons are mimicked by three spin-1/2 nuclei whose states are probed by another ancillary spin. It was also argued that the effect arises from the quantum contextuality in quantum physics. Later, Rae and Forgan describe, the effect is observed due to the interference between the wavefunctions of weakly interacting quantum particles \cite{RaeandForgan}. In order to close any conceptual loophole that may arise from the unknown local interactions of the physical qubits, Paraoanu illustrated the violation based on non-local schemes by designing two different quantum circuit using standard gates and measurement \cite{paraoanu2018nonlocal}. 

IBM’s cloud-based quantum computing platform has opened a new window of opportunity to perform experiments with quantum states. It allows testing various quantum mechanical phenomena \cite{GarciaJAMP2018,SisodiaQIP2017,HuffmanPRA2017,VishnuQIP2018,AlsinaPRA2016,YalcinkayaPRA2017,KandalaNAT2017,SisodiaPLA2017}. Here, we present an equivalent quantum circuit design using IBM’s real quantum processor ‘ibmqx2’ to investigate the quantum pigeonhole effect. In order to get rid of any kind of local interactions we implement two similar non local circuits proposed by Paraoanu \cite{paraoanu2018nonlocal}. We show that by standard quantum gate operations and measurements, it is indeed possible to observe quantum pigeonhole effect. We perform simulation to verify the theoretical predictions.

This paper is organized in the following way. In Section \ref{II}, the quantum pigeonhole effect is discussed briefly. In Section \ref{III}, we present the implementation of the quantum circuits on ‘ibmqx2’ superconducting chip to investigate the quantum pigeonhole effect and discuss the experimental outcome and its significance. In Section \ref{IV}, we give the conclusion about the work with some remarks.

\section{Theory \label{II}}
\begin{figure} [H]
    \centering
    \includegraphics [width=\linewidth]{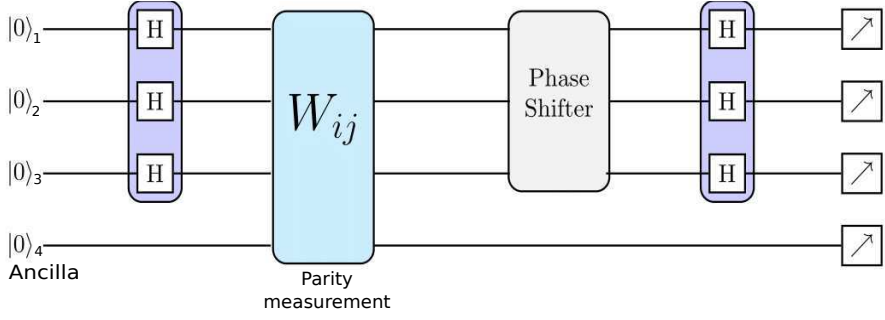}
    \caption{\textbf{The circuit schematic for investigation of the Quantum Pigeonhole Effect.} Hadamard gates are used to prepare the initial state of the qubits. The phase shifter perform the phase shift operation on the qubits. The 2-qubit parity measurement $W_{lm}$ is performed in order to retrieve the intermediate state information of the qubits. Another Hadamard operation is performed before the qubits are measured at the end of the circuit.}
    \label{fig2}
\end{figure}

In our experiment, we model the quantum pigeonhole effect using superconducting qubits in IBM quantum experience platform as shown in Figure \ref{fig2}). We consider a three qubit system which corresponds to three pigeons and two orthogonal states $\ket{0}$ and $\Ket{1}$, represents two boxes. We prepare the initial state by applying Hadamard gate on the three qubits 

\begin{equation}
    \Ket{\psi_i} =\Ket{+}_1 \Ket{+}_2 \Ket{+}_3 .
\end{equation} 

where, $\Ket{+}=\frac{\ket{0}+\Ket{1}}{\sqrt{2}}$ and the indices 1,2,3 refer to the qubits one, two and three respectively.

A phase-shifter is then operated on the initial state $\Ket{\psi_i}$ and the state transforms into $\Ket{+i}_1 \Ket{+i}_2 \Ket{+i}_3$ where, $\Ket{+i}=\frac{\ket{0}+i \Ket{1}}{\sqrt{2}}$. Then, after applying Hadamard gate, the three qubit state becomes 

\begin{align}
\Ket{\psi_f} &= \left( \frac{1+i}{2}\Ket{0} + \frac{1-i}{2}\Ket{1}  \right) \otimes \left( \frac{1+i}{2}\Ket{0} + \frac{1-i}{2}\Ket{1}  \right) \nonumber \\ 
& \hspace{2.9cm} \otimes \left( \frac{1+i}{2}\Ket{0} + \frac{1-i}{2}\Ket{1}  \right).
\end{align}

So, each qubit has equal probability to be found in either $\Ket{0}$ or $\Ket{1}$ state.

The $\Ket{+}$ state can also be written as

\begin{equation}
    \Ket{+}=\frac{1-i}{2} \Ket{+i} + \frac{1+i}{2} \Ket{-i}. 
\end{equation}

After the phase shift operator, $\Ket{+i}$ will transform to $\Ket{-}= \frac{\Ket{0} - \Ket{1}}{\sqrt{2}} $ and finally to $\Ket{1}$, after the Hadamard operation. Similarly, $\Ket{-i}$ will transform to $\Ket{+}$ and then to $\Ket{0}$, after the Hadamard operation. From this we can infer that after the measurement if we get $\Ket{0}$, then it corresponds to a post-selected state $\Ket{-i} = \frac{\ket{0}-i \Ket{1}}{\sqrt{2}}$ just before the phase-shift operator. In the same way $\Ket{1}$ will corresponds to the post-selected state $\Ket{+i}$.

In order to probe whether any two qubits are in the same quantum state or not, we need to perform a 2-qubit parity measurement $W_{lm} (l,m=1,2,3;l\neq m)$ of any two of the three qubits. The intermediate state of the qubits are measured through an ancilla qubit. For the intermediate situation thus it is convenient to define projection operators for various pairing combinations of three qubits.

\begin{align} 
\Pi_{12}= \ket{0}\bra{0}\otimes \Ket{0}\bra{0}\otimes \mathbbm{1} + \Ket{1}\bra{1}\otimes \Ket{1}\bra{1}\otimes \mathbbm{1} \nonumber \\ 
\Pi_{23}= \mathbbm{1} \otimes \Ket{0}\bra{0}\otimes \Ket{0}\bra{0}+ \mathbbm{1} \otimes \Ket{1}\bra{1}\otimes \Ket{1}\bra{1} \nonumber \\ 
\Pi_{13}=\Ket{0}\bra{0}\otimes \mathbbm{1} \otimes \Ket{0}\bra{0} + \Ket{1}\bra{1}\otimes \mathbbm{1} \otimes \Ket{1}\bra{1}.
\label{eqn4}
\end{align}

The projection operators in equation \ref{eqn4} tell us whether any two of the three qubits are in the same state or not. 
\begin{equation}
    \Pi_{lm}\Ket{\psi_i}=\Ket{\psi^{same} _{l,m}} \hspace{0.3cm} (l,m= 1,2,3 \hspace{0.1cm}; l \neq m).
\end{equation}

There are eight possible measurement outcome with equal probability 
\newline
\{$\Ket{000},
\Ket{001},\Ket{010},\Ket{011},\Ket{100},\Ket{101},\Ket{110},\Ket{111}$\} Now we would have expected from pigeonhole principle that at least two of the three qubits
take the same quantum state but we can see that for a particular instance where the measurement outcome is $\Ket{000}$ (or $\Ket{111}$), which corresponds to the post-selected state $\Ket{-i-i-i}$ (or $\Ket{+i+i+i}$) $\ket{\psi^{same} _{l,m}}$ is orthogonal to the post selected state $\Ket{-i-i-i}$ (and $\Ket{+i+i+i}$) prior to the phase-shift operator.

\begin{align}
    \braket{{-i-i-i}|{\psi^{same} _{l,m}}} = 0 \nonumber \\
    \braket{{+i+i+i}|{\psi^{same} _{l,m}}} = 0,
\end{align}
So we can infer that no two of the three qubits are found in identical states for the given particular post selected states $\Ket{-i-i-i}$ and $\Ket{+i+i+i}$. This result can be interpreted as Quantum Pigeonhole Effect (QPHE).
\\

\section{Circuit implementation and Results \label{III}}

\begin{figure}[]
    \centering
    \includegraphics [width=\linewidth]{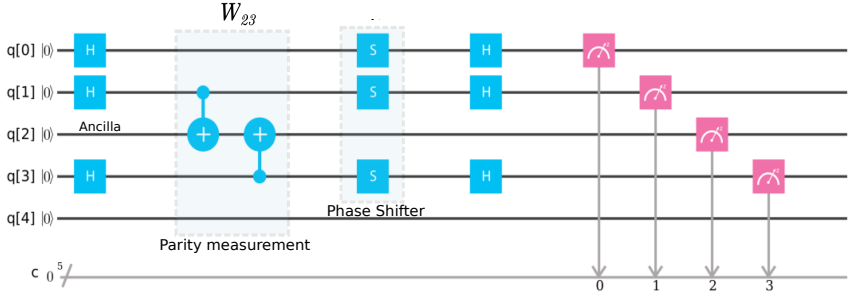}
    \caption{\textbf{The circuit implementation in IBM qauntum computer to investigate QPHE.} Hadamard gate is used to prepare the initial state. The parity measurement of the second (q[1]) and third(q[3]) qubit is performed using two consequtive CNOT gates with second (q[1]) and third (q[3]) qubit as control qubits and the ancilla qubit (q[2]) as the common target qubit. The phase gate $S$ introduce a phase shift. Hadamard operations is performed before all the qubits are measured at the end. }
    \label{fig3}
\end{figure}

For the experimental realization of Quantum Pigeonhole Effect we have used IBM's 5 qubit quantum computing interface ibmqx2. The circuit implementation for QPHE on ibmqx2 is shown in Figure \ref{fig3}. Here q[0], q[1] and q[3] are the three superconducting qubits. The information about the state of any qubit can be measured by using an ancilla q[2], another superconducting qubit.

\begin{table}[H]
    \centering
\begin{tabular}{|c|c|c|c|}
\hline
         \multicolumn{1}{|p{2cm}|}{\centering Post  \\ selected \\ state} &  \multicolumn{1}{|p{2cm}|}{\centering Parity  \\ measurement \\ $W_{1,2}$} & \multicolumn{1}{|p{2cm}|}{\centering Parity  \\ measurement \\ $W_{2,3}$} & \multicolumn{1}{|p{2cm}|}{\centering Parity  \\ measurement \\ $W_{1,3}$}  \\
         \hline
         $\Ket{{+i +i +i}}$ & 1 & 1 & 1 \\
         $\Ket{+i +i -i}$ & 1 & 0 & 0 \\
         $\Ket{{+i -i +i}}$ & 0 & 0 & 1 \\
         $\Ket{{+i -i -i}}$ & 0 & 1 & 0 \\
         $\Ket{{-i +i +i}}$ & 0 & 1 & 0 \\
         $\Ket{{-i +i -i}}$ & 0 & 0 & 1 \\
         $\Ket{{-i -i +i}}$ & 1 & 0 & 0 \\
         $\Ket{{-i -i -i}}$ & 1 & 1 & 1 \\
         \hline
\end{tabular} 
    \caption{2-qubit parity measurement for all possible post-selected state. $W_{lm}$ represents the 2-qubit parity measurement on $l th$ and $m th$ qubits. For the post selected state $\Ket{+i+i+i}$ (or $\Ket{-i-i-i}$) no two qubits are found in same state.}
        \label{tab1}
\end{table}

To probe the intermediate state information of any two ($(l,m= q[0],q[1],q[3]; l\neq m)$) of the three qubit prior to the phase-shifter, we perform a 2-qubit parity measurement $W_{lm}$ using a pair of CNOT gates ($C_l NOT_{q[2]},C_m NOT_{q[2]}$) where the ancilla q[2] acts as a common target bit. The parity measurement operator $W_{lm}$ preserve the state of the ancilla if both $l th$ and $m th$ qubits are in the same quantum state and inverts otherwise.

The intermediate information about the states of any pair ($l,m= 1,2,3 \hspace{0.1cm}; l \neq m$) of qubits can be obtained by measuring the state of ancilla. The state of the ancilla $0$ indicates that the pair of qubits are in the same state and the state of the ancilla $1$ corresponds to the situation where the two qubits are in different state. Table \ref{tab1} shows the results of the outcome of the parity measurement for all possible post-selected states. For the post-selected states $\Ket{+i+i+i}$ and $\Ket{-i-i-i}$ we can see that no pair of qubits are in the same state, thus shows QPHE. In the remaining cases we can also observe some interesting effects, e.g for the post-selected state $\Ket{-i+i-i}$ we find that the qubit 1 and 2 are in same state, qubit 2 and 3 are in same state but qubit 1 and 3 are in different state. Similar effects are observed for all the other cases.  

\begin{figure} [H]
    \centering
    \includegraphics [width=\linewidth] {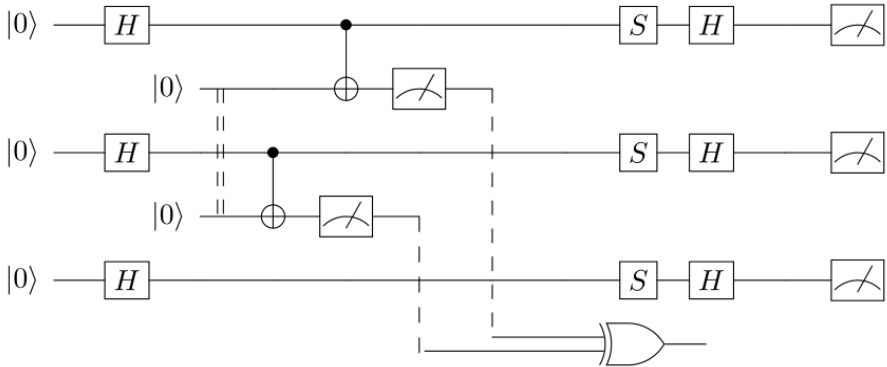}
    \caption{\textbf{The circuit illustrating QPHE based on entanglement distillation.} Two ancilla qubits are used for measuring the parity of the qubits, which are initially prepared in an entangled state $\frac{\Ket{00}+\Ket{11}}{\sqrt{2}}$. Here, the double dotted line represents the entanglement between the qubits and the single dotted line represents the classical channel.} 
    \label{fig4}
\end{figure}

The qubits might get disturbed due to some local interactions\cite{locality,realism} or direct physical interactions between qubits while performing the parity measurement using CNOT gates which can change the pre-existing values of the qubits. To eliminate such local interaction between the qubits, we consider two non-local set-ups as shown in Figure \ref{fig4} and Figure \ref{fig6}.

\begin{figure} [H]
    \centering
    \includegraphics [width=\linewidth]{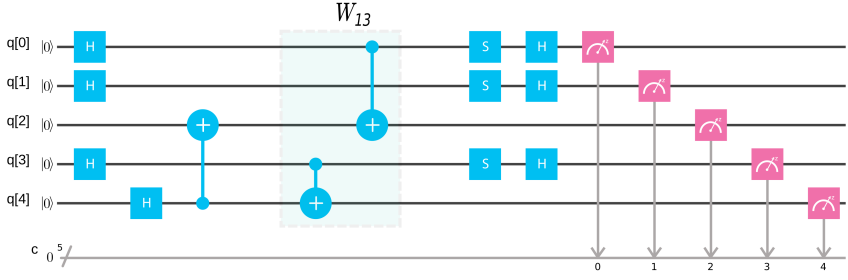}
    \caption{\textbf{The circuit implementation for QPHE based on entanglement distillation in IBM quantum computer.} Here, q[0], q[1], and q[3] are considered as system qubits, q[2] and q[4] are used as ancilla qubits for parity measurement.}
    \label{fig5}
\end{figure}

\begin{table}[H]
    \centering
\begin{tabular}{|c|c|c|c|}
\hline
         \multicolumn{1}{|p{2cm}|}{\centering Post  \\ selected \\ state} &  \multicolumn{1}{|p{2cm}|}{\centering Parity  \\ measurement \\ $W_{1,2}$} & \multicolumn{1}{|p{2cm}|}{\centering Parity  \\ measurement \\ $W_{2,3}$} & \multicolumn{1}{|p{2cm}|}{\centering Parity  \\ measurement \\ $W_{1,3}$}  \\
         \hline
         $\Ket{{+i +i +i}}$ & 1 & 1 & 1 \\
         $\Ket{+i+i-i}$ & 1 & 0 & 0 \\
         $\Ket{{+i -i +i}}$ & 0 & 0 & 1 \\
         $\Ket{{+i -i -i}}$ & 0 & 1 & 0 \\
         $\Ket{{-i +i +i}}$ & 0 & 1 & 0 \\
         $\Ket{{-i +i -i}}$ & 0 & 0 & 1 \\
         $\Ket{{-i -i +i}}$ & 1 & 0 & 0 \\
         $\Ket{{-i -i -i}}$ & 1 & 1 & 1 \\
         \hline
         \end{tabular}
    \caption{2-qubit non-local parity measurement based on entanglement distillation for all possible post-selected state. $W_{lm}$ represents the 2-qubit non-local parity measurement on $l th$ and $m th$ qubits. For the post selected state $\Ket{+i+i+i}$ (or $\Ket{-i-i-i}$) no two qubits are in same state.}
    \label{tab2}
\end{table}

In the scheme shown in Figure \ref{fig4} the parity measurement is realized by two local CNOT gate operations which are converted into a classical parity assessment using classical XOR gate \cite{EntgDistil}. Here two ancilla qubits are entangled in the  $\Ket{\Phi^+}=\frac{\ket{00}+\Ket{11}}{\sqrt{2}}$ Bell state which act as the target qubits while performing the parity measurement using CNOT gates. The ancilla qubits are then measured and the measurement outcome is transmitted to a XOR gate as classical bits. If the output of the XOR gate is 0 (1) then it indicates that the two qubits are in the same (different) quantum state. The IBM circuit implementation is shown in Figure \ref{fig5}

In another possible non-local scheme, shown in Figure \ref{fig6} which is based on the idea of teleportation of CNOT gates \cite{TeleportCnot}. In the previous setup two entangled ancilla are used for measuring the parity. Now it is possible that some unknown effects from the first qubit which can propagate to the ancilla, and then to the second qubit. To avoid this, two ancilla are used for measuring the parity and another ancilla as their common target. If the  outcome of third ancilla is $0 (1)$, it corresponds that the qubits are in the same (different) state. The IBM circuit implementation is demonstrated in Figure \ref{fig7}
\begin{figure} [H]
    \centering
    \includegraphics [width=\linewidth]{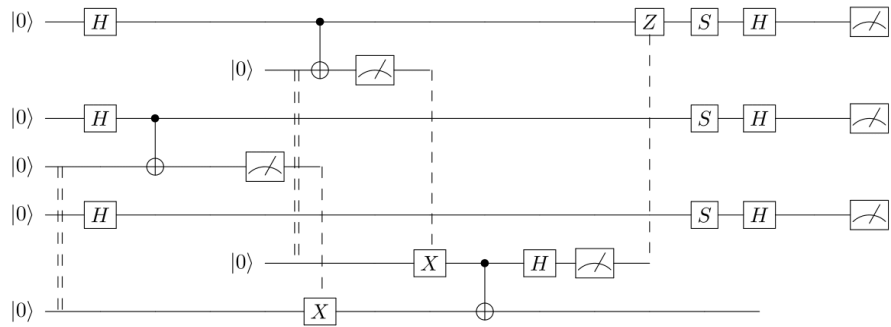}
    \caption{\textbf{The circuit illustrating QPHE based on teleportation of CNOT gates.} The double dotted line represents the entanglement between the qubits, and the single dashed line indicates classical communication. The measurement outcome of ancilla, 0 (or 1) corresponds to qubits being in the same state (or different state).} 
    \label{fig6}
\end{figure}

\begin{figure} [H]
    \centering
    \includegraphics [width=\linewidth]{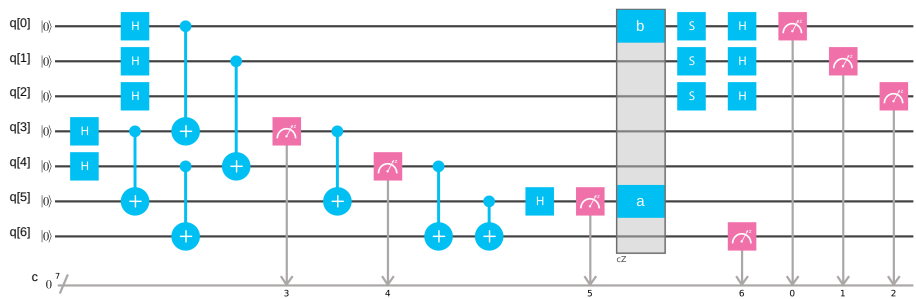}
    \caption{\textbf{The circuit implementation for QPHE based on teleportation of CNOT gates in IBM quantum computer.} q[0], q[1] and q[2] are the system qubits and q[3], q[4], q[5] and q[6] are the ancilla qubits.}
    \label{fig7}
\end{figure}

\begin{table}[H]
    \centering
\begin{tabular}{|c|c|c|c|}
\hline
         \multicolumn{1}{|p{2cm}|}{\centering Post  \\ selected \\ state} &  \multicolumn{1}{|p{2cm}|}{\centering Parity  \\ measurement \\ $W_{1,2}$} & \multicolumn{1}{|p{2cm}|}{\centering Parity  \\ measurement \\ $W_{2,3}$} & \multicolumn{1}{|p{2cm}|}{\centering Parity  \\ measurement \\ $W_{1,3}$}  \\
         \hline
         $\Ket{{+i +i +i}}$ & 1 & 1 & 1\\
         $\Ket{+i+i-i}$ & 1 & 0 & 0 \\
         $\Ket{{+i -i +i}}$ & 0 & 0 & 1 \\
         $\Ket{{+i -i -i}}$ & 0 & 1 & 0 \\
         $\Ket{{-i +i +i}}$ & 0 & 1 & 0 \\
         $\Ket{{-i +i -i}}$ & 0 & 0 & 1 \\
         $\Ket{{-i -i +i}}$ & 1 & 0 & 0 \\
         $\Ket{{-i -i -i}}$ & 1 & 1 & 1 \\
         \hline
         \end{tabular}
        \caption{2-qubit non-local parity measurement based on teleportation of CNOT gates for all possible post-selected state. $W_{lm}$ represents the 2-qubit non-local parity measurement on $l th$ and $m th$ qubits. For the post selected state $\Ket{+i+i+i}$ (or $\Ket{-i-i-i}$) no two qubits are in same state}
    \label{tab3}
\end{table}

Table \ref{tab2} and \ref{tab3} shows the results obtained using aforementioned two non-local schemes based on entanglement distillation and teleportation of CNOT gates respectively. From both of the table \ref{tab2} and \ref{tab3} it is pretty evident that for the post-selected state $\Ket{+i+i+i}$ and $\Ket{-i-i-i}$ the outcome of the non-local parity measurement indicate that none of the two qubits are in the same quantum state and it also explins the fact that the effect is non-local. For other post-selected states the results resembles similar patterns as we have observed in \ref{tab1}. Overall, we can see a consensus among the results from  all the three schemes which match well with the theoretical predictions of the quantum pigeonhole effect.
\\

\section{Conclusion \label{IV}}
To summarize, we have designed and successfully implemented a suitable quantum circuit to observe the quantum pigeonhole effect. We have performed experimental simulation using IBM’s real quantum processor ‘ibmqx2’. Each qubit is prepared and post-selected individually. If we measure the state of each qubit separately, they appear to be completely uncorrelated, but when we make a joint measurement on the pairs of qubit we find them to be correlated. This correlation is a manifestation of the quantum pigeonhole effect. We have also shown that the correlation exist in a non-local set-up where any possible unknown local interactions is eliminated using non-local parity measurement technique. 

\section*{Acknowledgements}
The authors acknowledge the use of IBM Quantum Experience platform for this work and grateful to IBM team. The discussions and opinions developed in this paper are only those of the authors and do not reflect the opinions of IBM or IBM Q experience team.

\section*{Author contributions}
The first three authors NNH, AD, SS contributed equally to the work. NNH designed the quantum circuit and implemented on the IBM’s quantum processor along with AD and SS. NNH, AD and SS performed experimental simulations and completed the work under the guidance of PKP. All authors discussed the results and contributed to the final manuscript.

\end{document}